\documentclass[twocolumn,showpacs,preprintnumbers,amsmath,amssymb,pra]{revtex4}
\usepackage{graphicx}
\usepackage{dcolumn}
\usepackage{bm}

\begin{document}

\title{1D to 3D beyond-mean-field dimensional crossover in mixture quantum droplets}
\author{L. Lavoine and T. Bourdel
\footnote{Email: thomas.bourdel@institutoptique.fr}
}
\affiliation{Laboratoire Charles Fabry, UMR 8501, Institut d'Optique, CNRS, Univ Paris-Saclay, Avenue Augustin Fresnel, 91127 PALAISEAU CEDEX}
\date{\today}
\begin{abstract}
The existence of quantum droplets in binary Bose-Einstein condensate mixtures rely on beyond-mean field effects, competing with mean-field effects. Interestingly, the beyond-mean field effect is changing from repulsive in 3D to attractive in 1D leading to drastically different behaviors. We study in detail the crossover between these two regimes in a quasi-1D regime where the radial wave-function is frozen. Quantum droplets exist for all values and sign of the mean-field interaction. We find that approaching the crossover is experimentally appealing as it reduces the relative importance of three-body losses and give realistic numbers for the realization of quantum droplet in the crossover, which would permit to test beyond-mean-field theories to an unprecedented precision. 
\end{abstract}

\pacs{
} 

\maketitle

\section{Introduction}
Ultracold quantum gases are unique well controlled many-body systems \cite{Bloch2012}. Their diluteness permits accurate ab initio theoretical treatment using zero range interaction. They are thus good candidates for tests of many-body theories. As an exemple, the energy of a zero temperature Bose gas can be calculated within the Bogoliubov theory beyond the usual mean-field approximation \cite{Lee1957a, Stringari}. Experimentally, there has been a quest for the measurement of these beyond mean-field corrections mostly by increasing $n a^3$, where $n$ is the density and $a$ the scattering length \cite{Papp2008, Shin2008, Navon2011, Lopes2018}.

Recently, it was discovered that quantum mixtures of two Bose-Einstein condensates with repulsive intraspecies interaction and attractive interspecies interactions permit a cancellation of the global mean field interaction without a reduction of the magnitude of the beyond mean-field effects \cite{Petrov2015}. They can then play a dominant role in the dynamics of the system and compete with the reduced mean-field energy. In this context, quantum droplets, $i.e.$ self-bound Bose-Einstein condensates due to beyond-mean field effects have been predicted \cite{Petrov2015} and experimentally observed \cite{Cabrera2018, Semeghini2018, Cheiney2018, Derrico2019}. The name droplet is given in analogy to liquid droplets which have similar properties such as a constant density profile although their stabilization mechanism is different. Interestingly, quantum droplets were also observed in dipolar condensates where the magnetic interaction competes with the usual contact interaction \cite{Ferrier2016} (see \cite{Luo2020, Bottcher2020} and references therein for recent reviews on quantum droplets). Experimentally, the droplets are observed at high densities (typically $\sim 10^{21}$m$^{-3}$). Three-body losses thus play an important role in the droplet dynamics \cite{Ferioli2020} and have hindered the observation of a stable flat-top density profile. 

Interestingly, the beyond-mean field energy, which originates from the summation of the zero point energies of the Bogoliubov modes in the Lee-Huang-Yang description \cite{Lee1957}, strongly depends on the dimension of the systems with important consequences \cite{Petrov2016}. For example, the 3D beyond-mean field energy density is positive and scales as $n^{5/2}$, whereas the 1D beyond-mean field energy density is negative and scales as $n^{3/2}$. In the 1D case, a dominant beyond-mean field energy is obtained at low density in contrast to the 3D case. Quantum droplets thus exists in both cases in however quite different conditions, requiring in particular an opposite sign of the mean-field interaction \cite{Astrakharchik2018}. 

Experimentally, droplets in Bose-Bose mixtures have been observed not only in free-space \cite{Semeghini2018} but also in cigar and pancake traps corresponding to quasi-1D \cite{Cheiney2018} and quasi-2D situations \cite{Cabrera2018}, where the motion of particules is frozen in one or two directions. Nevertheless, the beyond-mean field effects were still in a 3D regime. This possibility of hybrid dimension comes from the different energy scales associated with the two excitation branches, which are relevant in mixture droplets, $i.e.$ the low energy density branch (the two condensates oscillate in phase) and the high energy spin branch (the two condensates oscillate out of phase), which is responsible for the main beyond-mean field effects. Going toward the pure 1D regime is appealing as the droplets have a lower density reducing the nuisance of three-body losses \cite{Zin2018}. 

In this paper, we study the 1D to 3D dimensional crossover for quasi-1D quantum droplets in the experimentally relevant case of a cigar-shaped harmonic trap. Our work crucially uses a calculation of the beyond-mean field energy density in a quasi-1D Bose gas, where the radial wave-function is fixed to the radial harmonic oscillator one \cite{Ilg2018}. We find that quasi-1D quantum droplets exist for any values (either positive or negative) of the mean-field term and that there is a smooth 1D-3D crossover where the spin mode excitations and thus also the Lee-Huang-Yang energy density are modified by the radial trapping. We take a special care in estimating realistic experimental parameters in this regime. The paper will first describe large bulk droplet properties and second introduce finite size effects within an extended Gross-Pitaevskii equation formalism. We will study the ground state droplets with and without longitudinal trapping as well as the breathing mode frequency as a function of the relevant dimensionless parameters. 

\section{Quantum droplets in the bulk}
We consider a mixture of two atomic Bose-Einstein condensates in states $|1\rangle$ and $|2\rangle$ (of equal mass $m$ for simplicity) and radially harmonically trapped with a frequency $\omega_\perp/2\pi$. We assume that we are in the quasi-1D regime such that the radial condensate wave-functions are gaussian characterized by the harmonic oscillator size $a_\perp=\sqrt{\hbar/m\omega_\perp}$, where $\hbar$ is the reduced Planck constant.  The three relevant scattering lengths are $a_{11}>0, a_{22}>0, a_{12}<0$ associated with the 1D coupling constants $g_{ij}=2\hbar \omega_\perp a_{ij}$. We first consider the homogenous case characterized with the two 1D densities $n_1$,  $n_2$ and the total density $n=n_1+n_2$.

The mean field energy density can be written as:
\begin{equation}
\begin{split}
E_{MF}&=g_{11}n_1^2/2+g_{22}n_1^2/2+g_{12}n_1n_2\\
E_{MF}&=\frac{(\sqrt{g_{11}}n_1-\sqrt{g_{22}}n_2)^2}{2}+\frac{g \delta g (\sqrt{g_{11}}n_2+\sqrt{g_{22}}n_1)^2}{(g_{11}+g_{22})^2}\\
&\text{with } \delta g=g_{12}+\sqrt{g_{11}g_{22}} \text{ and } g=\sqrt{g_{11}g_{22}} 
\end{split}
\end{equation}
In the vicinity of the mean-field collapse $\delta g/g \ll 1$, the first term is much larger than the second one. The system thus minimizes its mean-field energy by locking the two densities such that $\sqrt{g_{11}}n_1=\sqrt{g_{22}}n_2$. In this situation, the mean-field energy density reduces to
\begin{gather}
\dfrac{E_{MF}}{\hbar \omega_\perp}=\dfrac{2a\delta a}{(\sqrt{a_{11}}+\sqrt{a_{22}})^{2}}n^{2}\\
\text{with } \delta a=a_{12}+\sqrt{a_{11}a_{22}} \text{ and } a=\sqrt{a_{11}a_{22}} 
\end{gather}
This equation can be written in a more convenient form 
\begin{gather}
\dfrac{E_{MF}}{\hbar \omega_\perp}=\delta a' \lambda \kappa^{2}/2a
\end{gather}
with $\kappa=na$, $\lambda=a/a_\perp$, and  $\delta a'=\frac{4 \delta a}{\lambda (a_{11}^{1/2}+a_{22}^{1/2})^2}$ a dimensionless parameter characterizing the mean-field interaction.

The beyond-mean field energy density has been calculated in two limits depending on the value of $\kappa$. For $\kappa \gg 1$ corresponding to large densities such that the spin healing length is smaller than $a_\perp$, one can make a local density approximation along the radial gaussian density profile and integrate the usual 3D expression of the beyond mean-field \cite{Ilg2018}:
\begin{gather}
\frac{E_{BMF}^{3D}}{\hbar \omega_\perp}=\frac{\lambda}{a} \frac{512}{75\pi}\kappa^{5/2}.
\end{gather}

In the opposite limit $\kappa \lesssim 1$, the summation over the radial oscillation modes has been performed in order to calculate the beyond-mean field energy density of a quasi-1D Bose-Einstein condensate in the 1D-3D crossover \cite{Ilg2018}. Interestingly, the result can also be used for the spin modes in a Bose-Bose mixture, giving the following beyond-mean field contribution. 
\begin{gather}
\frac{E_{BMF}^c}{\hbar \omega_\perp}=\frac{\lambda}{a} f(\kappa) \textrm{ , with }\\
f(\kappa)=\frac{C_{1D}^h}{\sqrt{2}}\kappa^2-\frac{4\sqrt{2}}{3 \pi}\kappa^{3/2}+\frac{4 \sqrt{2} \log(\frac{4}{3})}{\pi}\kappa^{5/2}+B_{1D}^h \kappa^3.
\end{gather}
The first term in $f(\kappa)$ is a mean-field correction due to the confinement-induced resonance ($C_{1D}^h\approx 1.4603$). The second term is the beyond mean-field contribution of a purely one-dimensional system \cite{Petrov2016}, which dominates at small $\kappa$. The third and four terms are corrections for higher values of $\kappa$ ($B_{1D}^h\approx 1.13$ \cite{misprint}).
For $\kappa \sim 1$, the beyond-mean field energy density cannot be written in a simple form but one can interpolate between the two previous expressions with a relatively good accuracy. 

\begin{figure}[htbp!]
\includegraphics[width=0.4\textwidth]{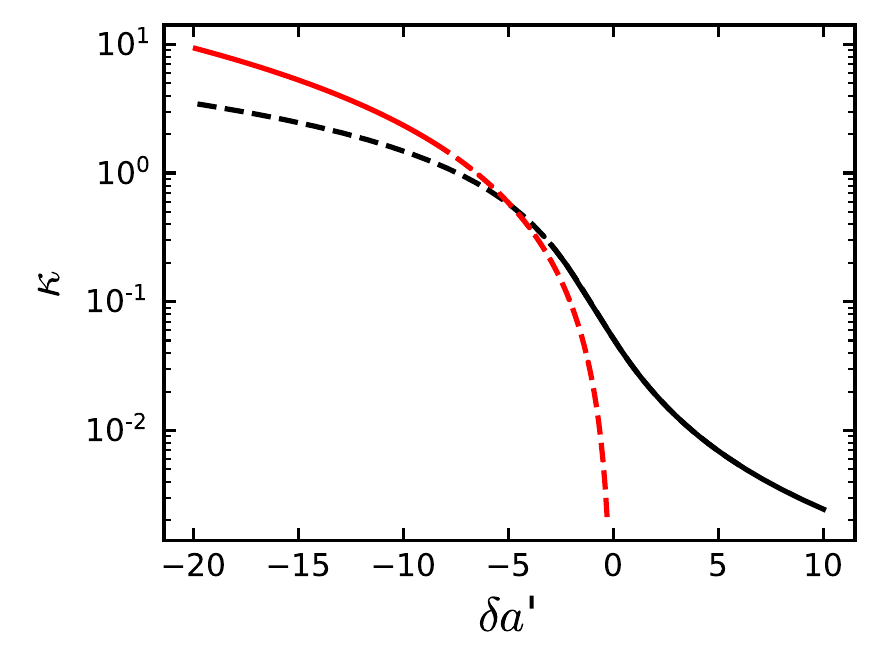}
\caption{(Color online) Dimensionless density $\kappa$ in a bulk quantum droplet as a function of the mean-field parameter $\delta a'$. The black (resp. red) curve corresponds to the 1D-3D crossover (resp. 3D) beyond-mean-field model. The curves are dashed in the regions where the theory is expected not to be valid.}
\label{kappa}
\end{figure}

In the two cases, one can simply minimize the energy per particules as a function of $\kappa$ in order to find the equilibrium density in the bulk, $i.e.$ neglecting the kinetic energy. We plot the resulting value of $\kappa$ as a function of the mean-field parameter $\delta a'$ (see Fig.\,\ref{kappa}).
Note that in the 3D beyond-mean field case, the minimization leads to a non-zero density only for attractive mean-field interaction which compensates a repulsive beyond mean-field term (red curve). On the contrary, the crossover expression leads to the existence of a finite density droplet for any value of the mean-field parameter (black curve). 
For large and negative mean-field parameter, we find $\kappa \gg 1$ and the 3D expression is the valid one. For $\delta a' \gtrsim -3$, the crossover expression finds $\kappa \lesssim 1$ and it is thus valid. By interpolating the two results from their validity region, we find droplets for any value of the mean-field interaction and an approximated value of $\kappa$ in the whole crossover. The equilibrium density is drastically reduced when increasing the mean-field interaction parameter. Going in this direction would greatly decrease the three-body loss rate that have been found to play an important role in previous experiments \cite{Ferioli2020}. 

 \begin{figure}[htbp!]
\includegraphics[width=0.4\textwidth]{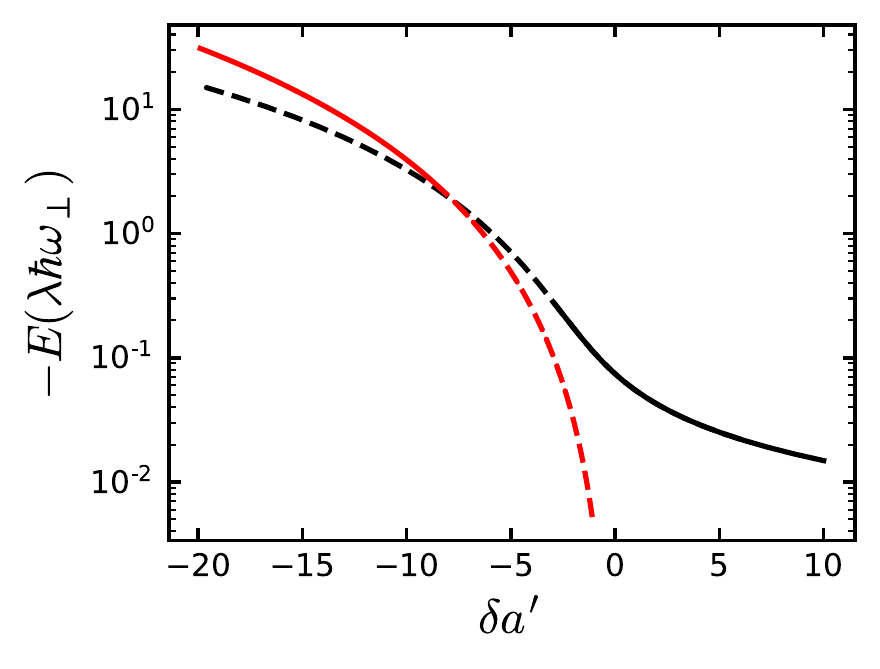}
\caption{(Color online) Energy per particle in the bulk. The black (resp. red) line corresponds to the energy minimization using the 1D-3D crossover (resp. 3D) formula. The curves are dashed in the regions where the corresponding theory is expected not to be valid.}
\label{VTharm}
\end{figure}
\begin{figure}[htbp!]
\includegraphics[width=0.4\textwidth]{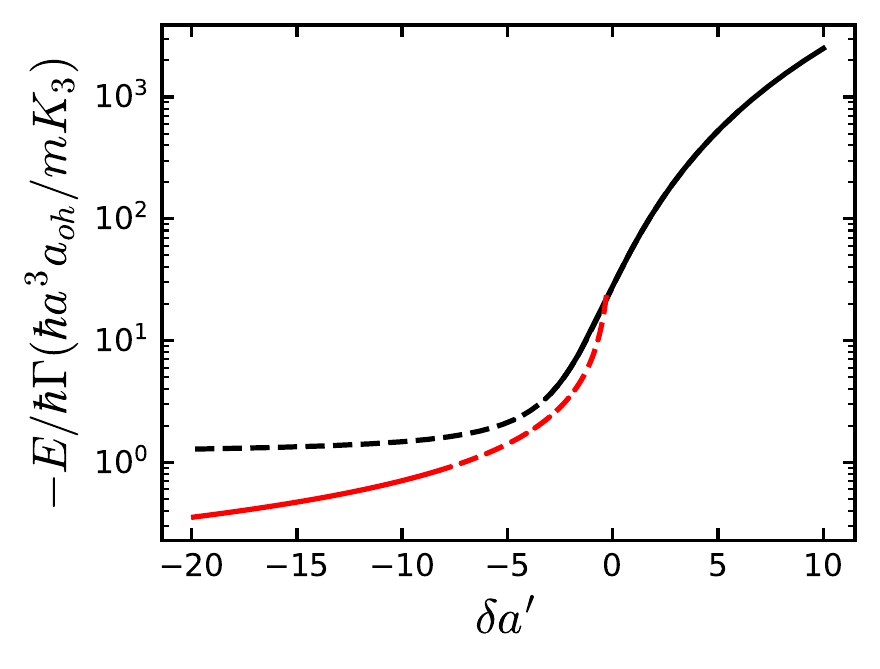}
\caption{(Color online) Ratio of the energy per particule to the three-body loss rate in the beyond-mean-field 1D-3D crossover. Larger ratios are favorable to minimize the relative effect of losses.}
\label{loss}
\end{figure}

Another interesting quantity is the energy per particle. It is negative as a droplet is a self bound object. One can interpolate between the two solid curves in order to find its behavior for any parameter $\delta a'$. It is clear that the binding energy of the droplet significantly decreases as one goes toward the 1D regime such that longer times will be necessary to observe it. In this context, we can thus wonder wether the increase lifetime of the droplet is sufficient to counterbalance its reduced energy scale. We thus plot the ratio of the droplet energy to the three-body loss rate. We find that it is indeed favorable to move toward $\delta a' >0$ in order to minimize the relative effect of losses (see Fig.\,\ref{loss}).

We now discuss experimentally realistic numbers. We take values of the scattering lengths for droplets made of potassium 39 in the second and third spin states around $56.7\,$G as these are the ones with appropriate signs and used in previous experiments. In this case $a_{11}\approx 33\,a_0$, $a_{22}\approx 84\,a_0$, where $a_0$ is the Bohr radius, and $\delta a$ and thus $\delta a'$ can be varied around zero by slight adjustment of the magnetic field. As an example, we take $\omega_\perp/2\pi=500\,$Hz and the energy scale $\lambda \hbar \omega_\perp \approx (2\pi \hbar)\times2\,$Hz is relatively low. For $\delta a=-5\,a_0$, a value close to the experimental ones \cite{Cabrera2018, Semeghini2018, Cheiney2018}, $\delta a'=-20$ and the binding energy per particle $\sim(2\pi \hbar)\times60 \,$Hz is such that droplet physics can be observed on a time scale of tens of milliseconds \cite{Cheiney2018}. With an average effective three-body loss rate coefficient $K_3=1.4\times10^{-40}\,$m$^6.$s$^{-1}$ \cite{Cheiney2018}, the loss rate is $\sim 80\,$s$^{-1}$ and it plays a significant role in the dynamics. 

When one goes toward the 1D regime, the energy scale decreases rapidly. Nevertheless, the loss rate decreases even more rapidly and the crossover regime could be an adequate region to look for more stable droplets. For example at $\delta a'=0$, where the droplet is solely stabilized by the peculiar density dependance of the beyond-mean field energy as a function of density \cite{Ilg2018}, the droplet density is reduced by a factor $\sim$100 and three-body losses are then negligible. The energy per particle is then of the order of $\sim(2\pi \hbar)\times0.2\,$Hz. Note that such a low energy scale implies long experimental times as well as a control of the trap parameters such as its longitudinal flatness to an extreme precision. 

\section{Quantum droplets in a finite system}
Whereas the above discussion focused on the properties of droplets in the bulk, we now turn to the question of finite atom numbers and finite sizes in quasi-1D droplets. 

\subsection{Extended Gross-Pitaevskii equation}

In the same spirit as before, we suppose that the spin modes are unpopulated such that the ratio of densities between the two spin-states is fixed. Within this assumption the system can be described by a single wave-function $\psi(x,t)$ solution of the following extended Gross-Pitaevskii equation: 
\begin{gather}
i \hbar \frac{\partial \psi}{\partial t}=-\frac{\hbar^2}{2m}\frac{\partial^2 \psi}{\partial x^2}+\hbar \omega_\perp \left(\delta a' \lambda |\psi^2|+\lambda g(\kappa)\right)\psi+\frac{1}{2}m \omega_\parallel^2 x^2 \psi
\end{gather}
where $g(\kappa)=\frac{\partial f(\kappa)}{\partial \kappa}$ is the beyond-mean-field interaction potential.

This equation can be written in dimensionless units using the following scalings
\begin{gather}
t=t_0 t'=\frac{1}{\lambda \omega_\perp}t'
\\
x=x_0 x'=\frac{a_{\perp}}{\lambda^{1/2}}x'
\\
\psi=\psi_0 \psi'=\frac{1}{a^{1/2}}\psi'
\\
i\frac{\partial \psi'}{\partial t'}=-\frac{1}{2}\frac{\partial^2 \psi'}{\partial x'^2}+\left(\delta a' |\psi'^2|+g(|\psi'^2|)\right)\psi'+\frac{1}{2} \frac{\omega_\parallel^2}{\omega_\perp^2 \lambda^2} x'^2 \psi'
\end{gather}
with the normalisation condition $\int  |\psi'^2| \text{d} x'=N'=N\lambda^{3/2}$. In addition to this rescaled atom number, there are two dimensionless parameters: $\delta a'$ for the mean-field interaction and $\frac{\omega_\parallel}{\omega_\perp \lambda}$ for the trapping potential. 

\subsection{Ground state solutions}

\begin{figure}[htbp!]
\includegraphics[width=0.35\textwidth]{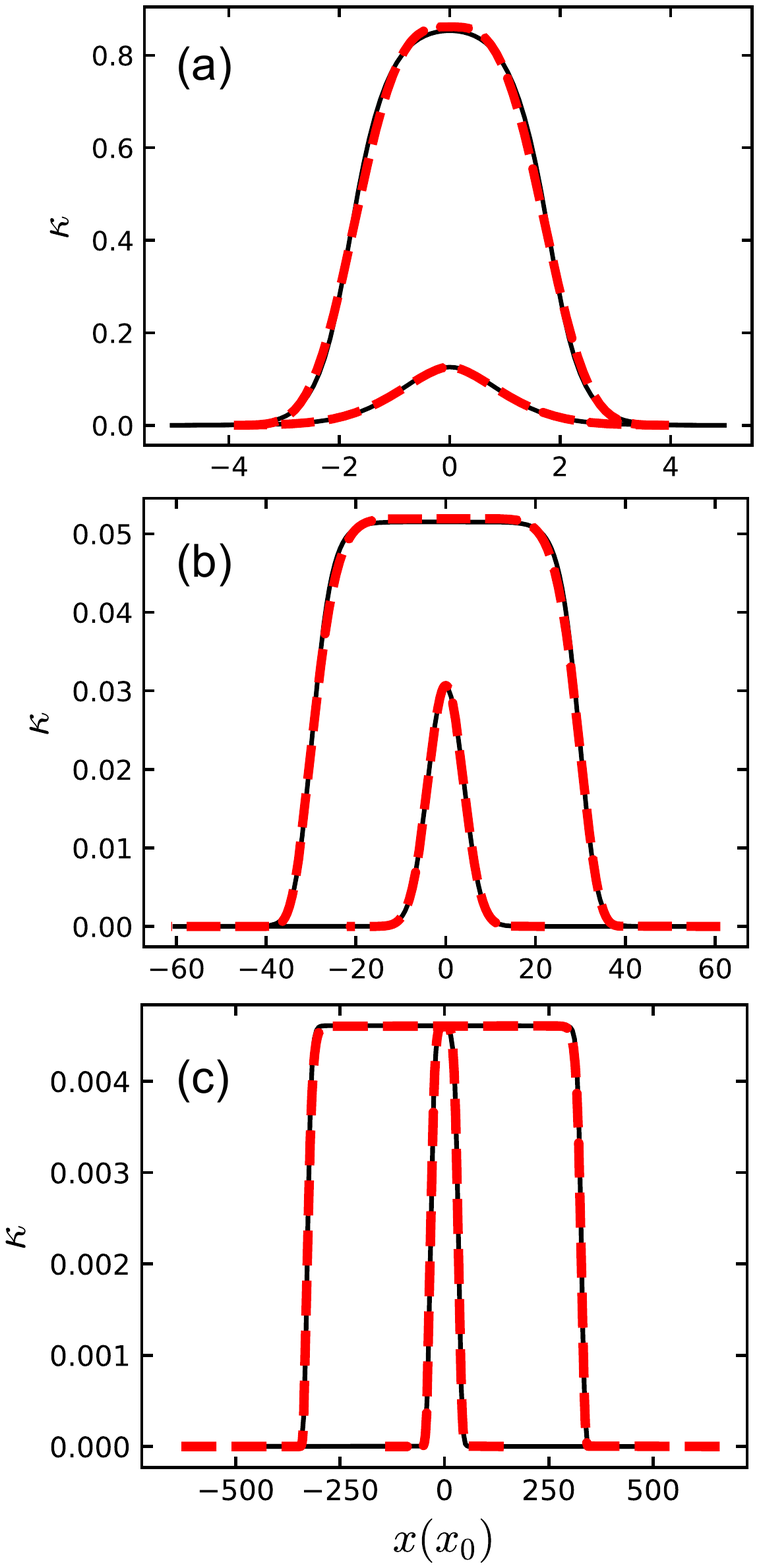}
\caption{(Color online) Quantum droplet density profiles for $N'=0.3$ and $N'=3$ for three values of $\delta a'$. (a): $\delta a'=-6.7$. (b): $\delta a'=0$.  (c): $\delta a'=6.7$. Please note, the different scales both in $x$ and in $\kappa$ in the three different figures. The solid black curves correspond to the exact minimization of the extended Gross-Pitaevskii equation in the 1D-3D crossover using imaginary time evolution. The superimposed dashed red curves are the results using the two parameter ansatz presented in the text.}
\label{4}
\end{figure}

We find the droplet ground state using the split step Fourier method and imaginary time evolution.  We first study the solution with no longitudinal trapping. Remarkably we always find a self bound solutions for all parameters. This is in contrast to the 3D situation where a minimal atom number is needed to form a droplet. This is also in contrast to the beyond mean-field 3D regime which needs $\delta a<0$ or the pure 1D regime which needs $\delta a>0$ for a self-bound solution to exist. We now plot the ground state density profiles for two different atom numbers $N'=0.3$ and $N'=3$ and mean-field interaction parameters in the crossover $\delta a'=$-6.7, 0, and 6.7 (corresponding to $\delta a/a_0=$-1.6, 0, and 1.6, for the previously given experimental parameters) in figure \ref{4}. 
For large atom numbers $N'\geq3$, the droplet profile exhibits a flat-top profile corresponding to the bulk solution whose edges are rounded because of the kinetic energy term. The droplet density gets smaller and its size larger as $\delta a'$ goes from negative to positive values. For small atom numbers $N'\leq0.3$, the droplets do not show a flat region, the kinetic energy is playing a dominant role. 

For the realistic experimental parameters chosen previously, $N'=1$ corresponds to 3800 atoms and  the quasi-1D situation offers the possibility to saturate a droplet with realistic small atom numbers. This is contrast to the quasi-2D and even more the 3D droplets where the critical atom number to reach a flat top can be too high especially for low values of $|\delta a|$, which are favorable to reduce the role of three-body losses. 

Another way to get an idea of the density profiles is to minimize the energy of an ansatz wavefunction. Here, an appropriate chose for the density profile is 
\begin{gather}
n(\sigma, r)=\frac{N}{2\sigma\Gamma(1+1/2r)}\exp(-(x/\sigma)^r) .
\end{gather}
This two-parameter function has the ability to interpolate between a peaked density profile for low values of $r$, to a gaussian for $r=2$ and to a flat-top density profile for $r\gg 2$. Its typical width is given by $\sigma$. It has the great advantage that the different energy terms are analytical and can be simply written in terms of the $\Gamma$ function. The energy minimisation is then straightforward. In figure \ref{4}, it is obvious that the ansatz minimization method is able to approximate the exact ground-state. 

\subsection{Maximum density and RMS size: scalings}
\begin{figure}[htbp!]
\includegraphics[width=0.4\textwidth]{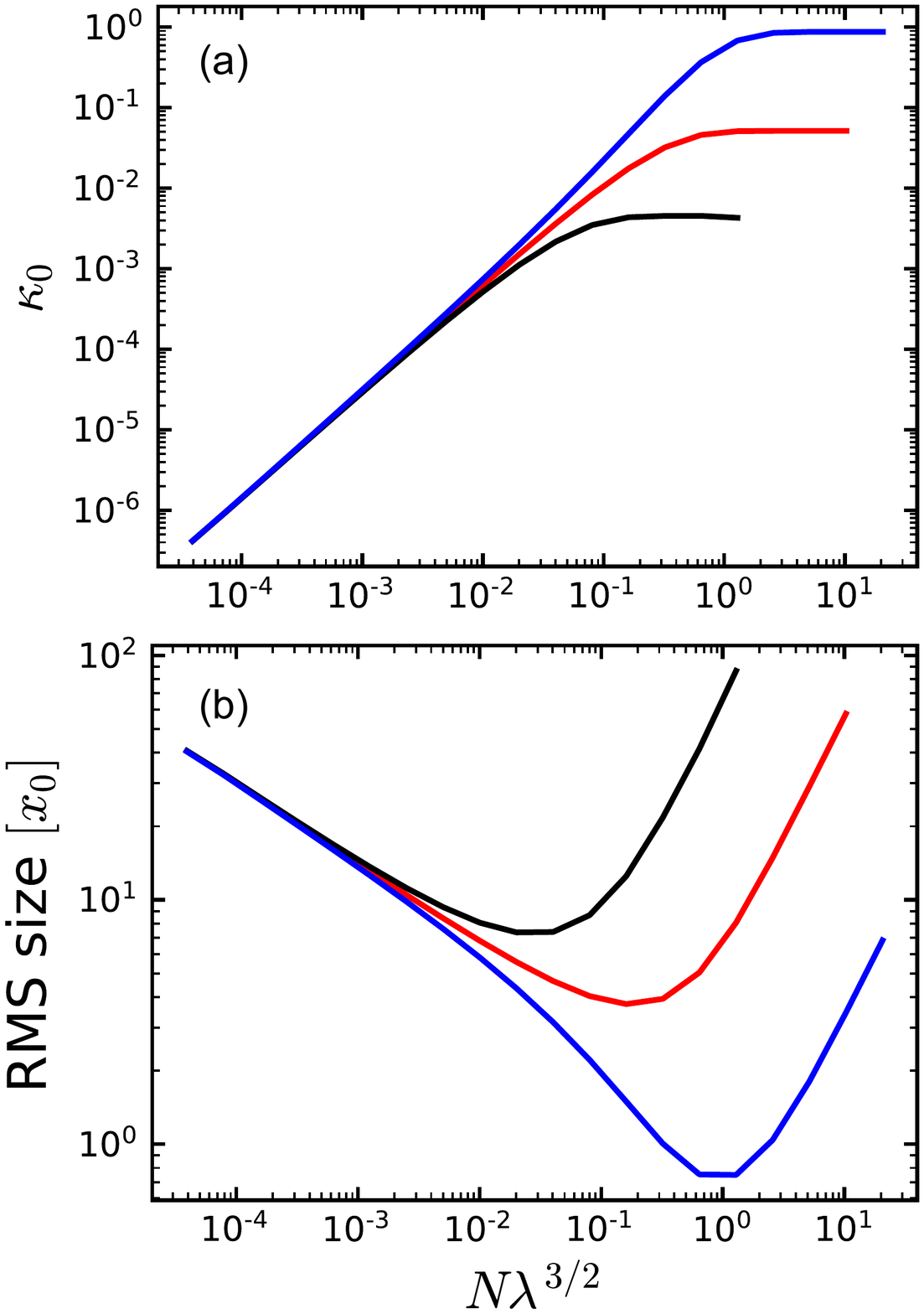}
\caption{(Color online) Maximum rescaled density $\kappa_0$ (a) and RMS size (b) of a droplet as a function of the atom number. The three curves correspond to three values of the mean-field parameters $\delta a'$. Black: $\delta a'=6.7$. Red (middle curve): $\delta a'=0$. Blue: $\delta a'=-6.7$.}
\label{RMS}
\end{figure}
We now plot the maximum rescaled density $\kappa_0$ and the root-mean-square (RMS) size of the ground-state profile as a function of the atom number for $\delta a'=$-6.7, 0, and 6.7 (see Fig.\,\ref{RMS}). In all cases, the density increases with the atom number until it reaches a saturation value corresponding to the bulk density, when the droplet exhibits a flat-top profile. It also appears that a lower atom number is necessary to reach a flat-top droplet as one moves toward the 1D regime. The ground-state size first decreases as the atom number increases and then increases when a flat-top droplet is formed. 

For the above figures, one can extract scalings in different regimes. For low atom numbers, all three curves are superimposed and $\kappa_0\ll1$. The dominant energy terms are the 1D beyond-mean-field attractive terms and the kinetic energy. The mean-field term is negligible. In this case, the size scales as $N'^{-1/3}$ and the density as $N'^{2/3}$. For large atom number, the ground-state is a flat-top quantum droplet. The dominant energies are then the mean-field and beyond mean-field terms. The kinetic energy can be neglected in the analysis. In this case, the size simply scales as $N'$ as the density is fixed. For attractive mean-field $\delta a'<0$, there is an intermediate situation corresponding to standard mean-field bright solitons \cite{Khaykovich2002, Strecker2002, Cheiney2018} for which the size scales as $N'^{-1}$ and the density as $N'^2$. 

\subsection{Trapped case}
\begin{figure}[htbp!]
\includegraphics[width=0.4\textwidth]{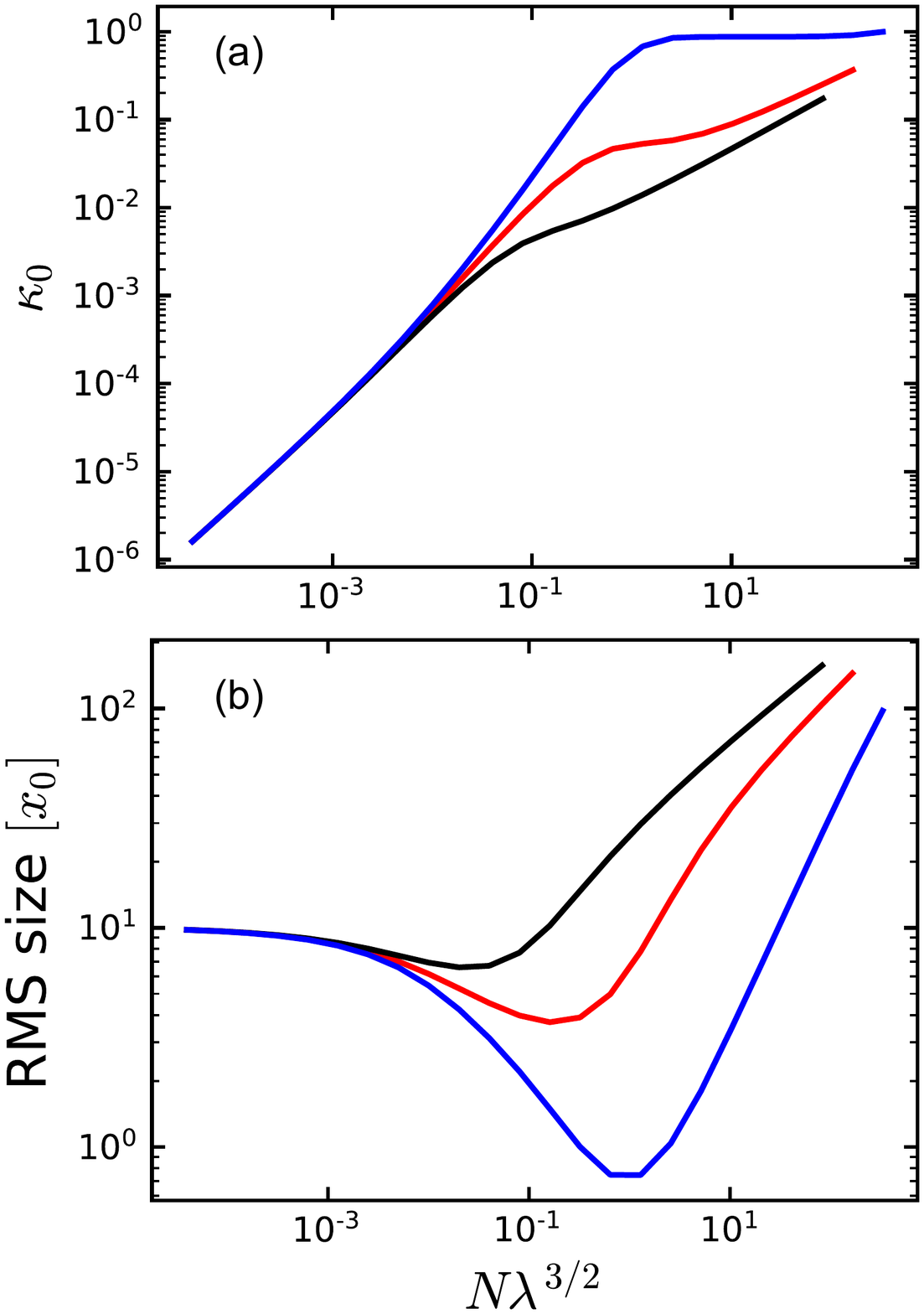}
\caption{(Color online) Maximum rescaled density $\kappa_0$ (a) and RMS size (b) as a function of the atom number in the presence of a longitudinal harmonic trap $\frac{\omega_\parallel}{\omega_\perp \lambda}=5\times 10^{-3}$. The three curves follow the same coding as in figure \ref{RMS}.}
\label{RMS_trap}
\end{figure}
We now turn to the trapped case. As an exemple, we chose $\frac{\omega_\parallel}{\omega_\perp \lambda}=5\times 10^{-3}$ (which corresponds to $w_\parallel/2\pi=0.01\,$Hz, for the previously chosen parameters) and find the ground state by imaginary time evolution (see Fig.\,\ref{RMS_trap}). For intermediate atom number, the trap has no effect. It corresponds to a regime where the trap potential energy is negligible as compared to the other energy scales. For low atom number, the size reaches a plateau in contrast to the untrapped case. This corresponds to a situation where the gas can be considered as non-interacting and the condensates occupy the ground state of the longitudinal harmonic trap. 

\begin{figure}[htbp!]
\includegraphics[width=0.35\textwidth]{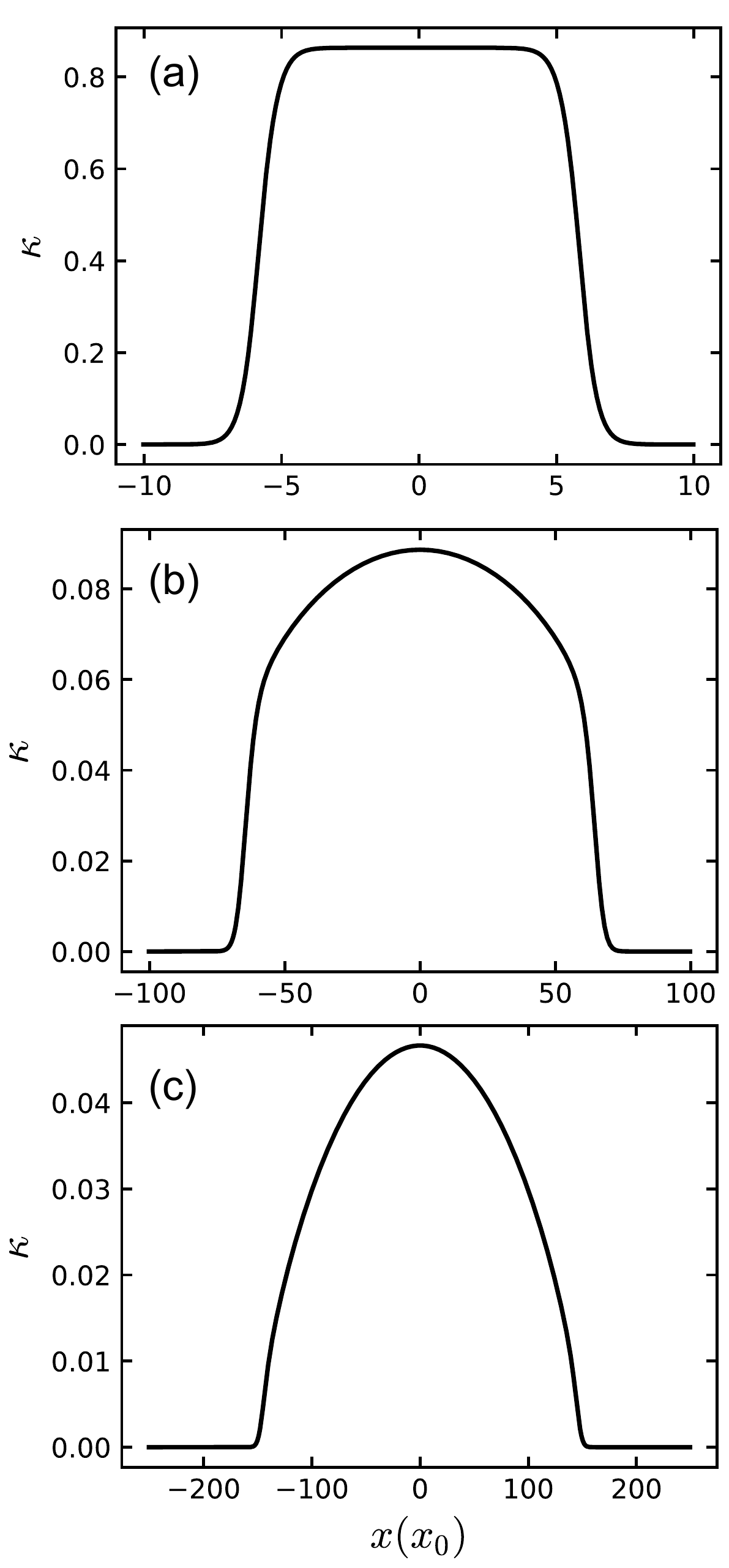}
\caption{(Color online) Droplet profiles for $N'=10$ for three different values of $\delta a'$. (a): $\delta a'=-6.7$. (b): $\delta a'=0$.  (c): $\delta a'=6.7$.}
\label{profils_trap}
\end{figure}

At high atom numbers, the trap prevents the droplet to grow in size at a contant density as observed in the absence of a longitudinal trap. In this regime, the kinetic energy can be neglected and the density profile can be found in an approximation analogous to the Thomas-Fermi approximation. The chemical potential is then directly linked to the density through the homogenous equation of state. In the 3D dimensional case, this was presented in \cite{Jorgensen2018}. The entrance in this last regime can be simply estimated by comparing the bulk energy per particule to the potential energy given the RMS size of the droplet. The value of $N'$ where this happens drastically depend on $\delta a'$. As a example, density profiles for $N'=10$ are plotted in figure \ref{profils_trap} for different values of $\delta a'$. The profile indeed ressembles a Thomas-Fermi profile for $\delta a'=6.7$ with dominant mean-field and potential energies. Oppositely, a flat-top droplet profile with negligible trap influence is found for $\delta a'=-6.7$. For $\delta a'=0$, the situation is intermediate. 

\subsection{Breathing mode}

\begin{figure}[htbp!]
\includegraphics[width=0.4\textwidth]{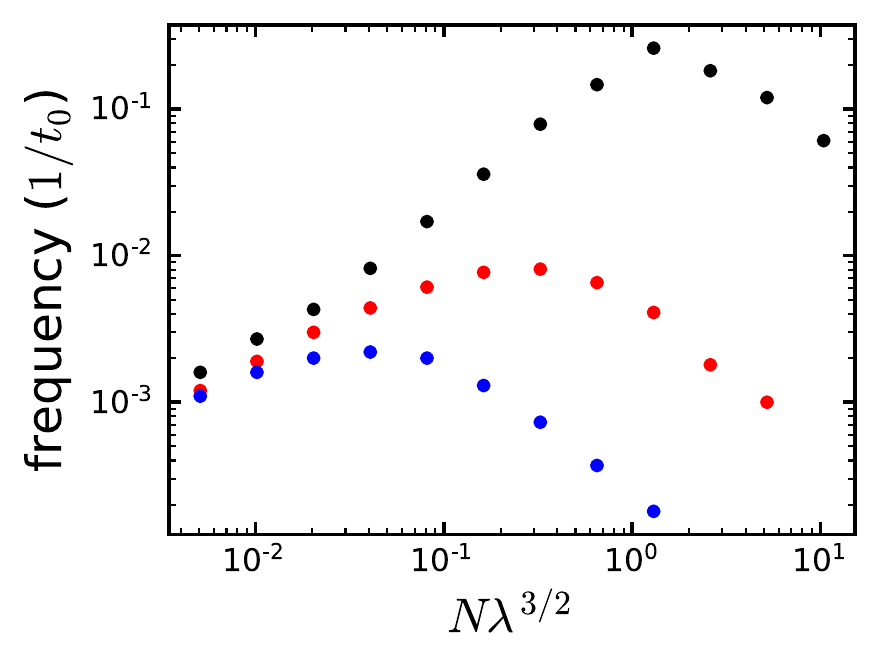}
\caption{(Color online) Frequency of the breathing mode as a function of the rescaled atom number for three values of the mean-field parameter $\delta a'$. Black top points: $\delta a'=-6.7$.  Red middle points $\delta a'=0$. Blue lower points:  $\delta a'=6.7$.}
\label{frequencies}
\end{figure}

We now turn to the study of excitations of the quasi-1D droplets in the absence of a longitudinal trap. They can be studied by real time integration of the extended Gross-Pitaevskii equation. More specifically, we start from a situation close from equilibrium by rescaling the first computed ground state wave-function by a coefficient of 1.01 \cite{Astrakharchik2018}. After this modification, we study in particular, the evolution of the RMS size of the droplet as a function of time and extract its main (lowest) oscillation frequency, which is plotted in figure \ref{frequencies}. Interestingly, we find that these small breathing oscillations are essentially undamped. This can be expected as their frequency remain below the particule emission threshold for all parameters. High frequency modes can evaporate but there are only little excited in our excitation scheme  \cite{Petrov2015}. This behavior is in contrast to the 3D situation where there is region just above the critical atom number where the droplets quickly evaporate to their ground-state \cite{Petrov2015}. Experimentally, excitations modes are important as they will be visible whenever the droplet is not prepared in a quasi-static way. 

We find that the breathing mode frequency is first increasing with the atom number as can be expected from an increasing binding energy. It then reaches a maximum close to the point where the droplet becomes flat top. Finally it decreases when the droplets get of larger size. This behavior observed in the whole crossover is similar to the one previously predicted in the pure 1D regime ($\delta a'>0$) \cite{Astrakharchik2018}. Note that the breathing oscillation frequencies are found to be small especially when increasing $\delta a'$ close to 0 or positive values. It will probably hinder the possibility of adiabatic preparation of ground state droplets in this regime. 

\section{Conclusions}
In mixture of two Bose-Einstein condensates with repulsive intraspecies interaction and attractive interspecies interaction, we have studied quantum droplets in the 1D-3D crossover for the beyond mean-field energy which changes positive in 3D to negative in 1D. We find that droplets exist for any value and sign of the total mean-field interaction. The equilibrium density is found to decrease rapidly when going toward the 1D regime, which would be experimentally favorable in order to reduce the role of three-body losses. This reduction comes together with a reduction of the typical energy scale of the droplets imposing severe contrains on the control of the residual potential such as the longitudinal trapping. Nevertheless, quantum droplets with a characteristic flat-top profile over a reasonably long life-time should be observable in the quasi-1D regime by reducing the value of $|\delta a|$ as compared to previous experiments \cite{Cabrera2018, Semeghini2018, Cheiney2018, Derrico2019}. The lifetime would be even longer for low loss mixtures such as Rb-K \cite{Derrico2019}.

Longer lifetimes would permit precise studies of the droplets properties and more generally of the beyond-mean-field effects in Bose gases. Effects beyond the standard Lee-Huang-Yang energy used in our work will appear, either because of higher order terms in the density expansion \cite{Ota2020}, because of finite-range interacting potentials \cite{Cikojevic2019, Cikojevic2020}, or because of temperature effects \cite{DeRosi2019}

\section{Acknowledgements}
We thank D. Petov for inspiring discussions and A. Hammond for his careful rereading. This  research  has  been  supported  by  CNRS,  Minst\`ere  de  l'Enseignement  Sup\'erieur  et  de  la  Recherche, Labex PALM, Region Ile-de-France  in  the  framework  of  DIM  Sirteq, Paris-Saclay in the framework of IQUPS, ANR Droplets - 19-CE30-0003-01, Simons foundation (award number 563916:  localization of waves).



\begin{thebibliography}{0}

\bibitem{Bloch2012}
I. Bloch, J. Dalibard, and S. Nascimb\`ene, Nature Physics {\bf 8}, 267 (2012).

\bibitem{Lee1957a}
T.D. Lee and C.N. Yang,  Phys. Rev. {\bf 105}, 1119 (1957).

\bibitem{Stringari}
L. Pitaevskii and S. Stringari, {\it Bose-Einstein Condensation and Superfluidity}, Oxford Science Publications. 



\bibitem{Papp2008}
S. B. Papp, J. M. Pino, R. J. Wild, S. Ronen, C. E. Wieman, D. S. Jin, and E. A. Cornell,
Phys. Rev. Lett. {\bf 101}, 135301 (2008).

\bibitem{Shin2008}
Y. I. Shin, A. Schirotzek, C. H. Schunck, and W. Ketterle, Realization of a Strongly,
Phys. Rev. Lett. {\bf 101}, 070404 (2008).

\bibitem{Navon2011}
N. Navon, S. Piatecki, K. G\"unter, B. Rem, T. C. Nguyen, F. Chevy, W. Krauth, and C.
Salomon, 
Phys. Rev. Lett. {\bf 107}, 135301 (2011).

\bibitem{Lopes2018}
R. Lopes, C. Eigen, N. Navon, D. Cl\'ement, R.P. Smith, and Z. Hadzibabic,
Phys. Rev. Lett. {\bf 119}, 190404 (2017).

\bibitem{Petrov2015}
D.S. Petrov, 
Phys. Rev. Lett. {\bf 115}, 155302 (2015).

\bibitem{Cabrera2018}
R. Cabrera, L. Tanzi, J. Sanz, B. Naylor, P. Thomas, P. Cheiney, and L. Tarruell,
Science  {\bf 359}, 301 (2018).

\bibitem{Semeghini2018}
G. Semeghini, G. Ferioli, L. Masi, C. Mazzinghi, L. Wolswijk, F. Minardi, M. Modugno, G. Modugno, M. Inguscio, and M. Fattori,
Phys. Rev. Lett.  {\bf 120}, 235301 (2018).

\bibitem{Cheiney2018}
P. Cheiney, C. R. Cabrera, J. Sanz, B. Naylor, L. Tanzi, and L. Tarruell, 
Phys. Rev. Lett.  {\bf 120}, 135301 (2018).

\bibitem{Derrico2019}
C. D'Errico, A. Burchianti, M. Prevedelli, L. Salasnich, F. Ancilotto, M. Modugno, F. Minardi, and C. Fort, 
Phys. Rev. Research  {\bf 1}, 033155 (2019).

\bibitem{Ferrier2016}
Igor Ferrier-Barbut, Holger Kadau, Matthias Schmitt, Matthias Wenzel, and Tilman Pfau,
Phys. Rev. Lett.  {\bf 116}, 215301 (2016).

\bibitem{Luo2020}
Zhihuan Luo, Wei Pang, Bin Liu, Yongyao Li, Boris A. Malomed, arXiv:2009.01061 (2020).

\bibitem{Bottcher2020}
F. B\"ottcher, J.-N. Schmidt, J. Hertkorn, K.S.H. Ng, S.D. Graham, M. Guo, T. Langen, T. Pfau, arXiv:2007.06391 (2020).

\bibitem{Ferioli2020}
G. Ferioli, G. Semeghini, S. Terradas-Brians\'o, L. Masi, M. Fattori, and M. Modugno,
Phys. Rev. Research  {\bf 2}, 013269 (2020).

\bibitem{Lee1957}
T.D. Lee, K.Huang and C.N. Yang, Phys. Rev. {\bf 106},1135 (1957).

\bibitem{Petrov2016}
D. S. Petrov and G. E. Astrakharchik, 
Phys. Rev. Lett.  {\bf 117}, 100401 (2016).

\bibitem{Astrakharchik2018}
G. E. Astrakharchik and B. A. Malomed, 
Phys. Rev. A  {\bf 98}, 013631 (2018).

\bibitem{Ilg2018}
 T. Ilg, J. Kumlin, L. Santos, and D. S. Petrov, and, H. P. B\"uchler, 
 Phys. Rev. A  {\bf 98}, 051604(R) (2018).

\bibitem{misprint}
There is a misprint in \cite{Ilg2018} for the value of $B_{1D}^h$. {\it Private communication}.

\bibitem{Zin2018}
P. Zin, M. Pylak, T.Wasak, M. Gajda, and Z. Idziaszek, 
Phys.Rev. A  {\bf 98}, 051603(R) (2018).

\bibitem{Khaykovich2002}
L. Khaykovich, F. Schreck, G. Ferrari, T. Bourdel, J. Cubizolles, L. D. Carr, Y.Castin, and C. Salomon, Science {\bf 296}, 1290 (2002).

\bibitem{Strecker2002}
K. E. Strecker, G. B. Patridge, A. G. Truscott, and R. G. Hulet, Nature {\bf 417}, 150 (2002).


\bibitem{Jorgensen2018}
N.B. J\/orgensen, G.M. Bruun, and J.J. Arlt
Phys. Rev. Lett.  {\bf 121}, 173403 (2018).

\bibitem{Bottcher2019}
F. B\"ottcher, M. Wenzel, J.-N. Schmidt, M. Guo, T. Langen, I. Ferrier-Barbut, T. Pfau, R. Bombin, J. S\'anchez-Baena, J. Boronat, and F. Mazzanti, Phys. Rev. Research  {\bf 1}, 033088 (2019).

\bibitem{Ota2020}
M. Ota, G.E. Astrakharchik, SciPost Phys. {\bf 9}, 020 (2020).



\bibitem{Cikojevic2019}
V. Cikojevi\'c, L.V. Marki\'c, G. E. Astrakharchik, and J. Boronat,
Phys. Rev. A 99, 023618 (2019).

\bibitem{Cikojevic2020}
V. Cikojevi\'c, L.V. Marki\'c, M. Pi, M. Barranco, and J. Boronat,
Phys. Rev. A 102, 033335 (2020).

\bibitem{DeRosi2019}
G. De Rosi, P. Massignan, M. Lewenstein, and G.E. Astrakharchik,
Phys. Rev. Research {\bf 1}, 033083 (2019).

\end{thebibliography}

\end{document}